\documentclass[reqno,a4paper,final]{amsart}
\usepackage{amssymb,amstext,amsthm,eucal}
\usepackage{bbold}
\usepackage{array}
\usepackage[latin1]{inputenc}
\newcommand{\half}{\tfrac{1}{2}}

\newcommand{\fd}{\mathfrak{d}}
\newcommand{\fg}{\mathfrak{g}}

\newcommand{\fso}{\mathfrak{so}}

\newcommand{\fsu}{\mathfrak{su}}

\newcommand{\SO}{\mathrm{SO}}

\newcommand{\Cl}{\mathrm{C}\ell}
\newcommand{\Spin}{\mathrm{Spin}}
\newcommand{\Sp}{\mathrm{Sp}}

\newcommand{\SU}{\mathrm{SU}}

\newcommand{\EE}{\mathbb{E}}
\newcommand{\RR}{\mathbb{R}}
\newcommand{\CC}{\mathbb{C}}

\newcommand{\VV}{\mathbb{V}}

\DeclareMathOperator{\CW}{CW}

\DeclareMathOperator{\AdS}{AdS}

\DeclareMathOperator{\dvol}{dvol}
\DeclareMathOperator{\grad}{grad}
\DeclareMathOperator{\Mat}{Mat}

\DeclareMathOperator*{\fS}{\mathfrak{S}}

\newcommand{\bv}{\boldsymbol{v}}
\newcommand{\bw}{\boldsymbol{w}}
\newcommand{\bx}{\boldsymbol{x}}

\newcommand{\blambda}{\boldsymbol{\lambda}}

\newcommand{\bH}{\boldsymbol{H}}
\newcommand{\bJ}{\boldsymbol{J}}

\newcommand{\1}{\mathbb{1}}

\newcommand{\repre}[1]{\underline{\mathbf{#1}}}
\allowdisplaybreaks[1]
\begin{document}
\title{On parallelisable NS-NS backgrounds}
\author{José Figueroa-O'Farrill}
\address{School of Mathematics, University of Edinburgh, Scotland, UK}
\email{j.m.figueroa@ed.ac.uk}
\thanks{EMPG-03-09}
%\date{\today}
\begin{abstract}
  We classify non-dilatonic NS-NS type II supergravity backgrounds
  admitting a consistent absolute parallelism.  They are all given by
  parallelised Lie groups admitting scalar flat bi-invariant
  lorentzian metrics.  There are seven different classes, some of them
  containing moduli.  For each class we determine the amount of
  supersymmetry which is preserved: there are examples with 16, 18,
  20, 22, 24, 28 and 32 supersymmetries.
\end{abstract}
\maketitle
\tableofcontents

\section{Introduction}

The purpose of this note is to present a classification of
non-dilatonic parallelisable NS-NS backgrounds of ten-dimensional type
II supergravity.  In this note, which has been prompted in part by the
preprints \cite{SSJ} and \cite{PandoZayas}, I attempt to place these
backgrounds in an appropriate mathematical context.  The main result
is that they are given by ten-dimensional Lie groups admitting a
bi-invariant scalar flat metric.  We then classify these groups up to
local isometry and explore the amount of supersymmetry that these
backgrounds preserve.  Let us start by introducing the context.

We will be dealing with the common sector of ten-dimensional type II
supergravity, the so-called NS-NS backgrounds where none of the RR
fluxes are turned on.  The data for such backgrounds consists of
\begin{itemize}
\item a ten-dimensional lorentzian spin manifold $(M,g)$,
\item a metric connection $D$ with \emph{closed} torsion three-form
  $H$, and
\item a dilaton $\phi$
\end{itemize}
subject to the equations of motion obtained by varying the action
functional, which takes the form
\begin{equation}
  \label{eq:action}
  \int_M e^{-2\phi} \left( R + 4 \|d\phi\|^2 - \half \|H\|^2 \right)
  \dvol_g ~.
\end{equation}
For the purposes of this note we will say that $(M,g)$ admits a
\emph{consistent absolute parallelism} (or is \emph{parallelisable},
for short) if $D$ is flat.  In this note we will classify the
parallelisable ten-dimensional spacetimes up to local isometry and
determine which of them are non-dilatonic supergravity backgrounds.

One can of course analyse a more general problem: namely classifying
those parallelised supergravity backgrounds for which the dilaton is
\emph{not} constant.  Besides the dilaton equation of motion, there is
another condition on a nonconstant dilaton.  Because the torsion
three-form $H$ is closed, $(M,g)$ is parallelisable if and only if it
is locally isometric to a Lie group admitting a bi-invariant metric.
In particular this means that $H$ is parallel, hence co-closed.  This
imposes constraints on the derivative of the dilaton.  From the
Maxwell equation for $H$, we see that
\begin{equation*}
  d\phi \wedge \star H = 0~,
\end{equation*}
which is equivalent to
\begin{equation*}
  \iota_{\grad \phi} H  = 0~,
\end{equation*}
so that relative to a basis of left-invariant vector fields, the
gradient of $\phi$ only has components along the centre of the Lie
algebra.  In other words, if $\theta$ denotes the left-invariant
Maurer--Cartan one-form then $\theta(\grad\phi)$ is a central element
of the Lie algebra.  We will \emph{assume} from now on that the
dilaton is constant, although this is not a consequence of
parallelisability.  In fact, since we will start by classifying the
parallelisable geometries, it is then a simple matter (which we will
nevertheless not address in this note) to determine which dilatons are
consistent with the parallelisable spaces.  This may in turn
constraint the geometry further and in any case will alter the
analysis of the supersymmetry of the backgrounds, to which we know
turn.

An important invariant of a supergravity background is the amount of
supersymmetry that it preserves, measured by the dimension of the
(linear) space of Killing spinors.  In principle this could be any
integer from $0$ to $32$, but it will be severely constrained in the
backgrounds under consideration.  In type II supergravity, there are
two types of Killing spinor equation, resulting from the supersymmetry
variations of the gravitino and of the dilatino.  The gravitino
variation gives a differential equation which says that a Killing
spinor is covariantly constant with respect to the spin connection
associated with $D$.  Since $D$ is flat for a parallelisable manifold,
this condition does not reduce the amount of supersymmetry of the
background, a fact also observed in \cite{SSJ}.  In contrast, the
dilatino variation gives an algebraic condition which says that a
Killing spinor is annihilated under Clifford multiplication by the
torsion three-form $H$ (for constant dilaton).  For a Lie group
admitting a bi-invariant lorentzian metric, $H$ is given essentially
by the structure constants relative to a frame consisting of
left-invariant vector fields.

The rest of the paper is organised as follows.  In
Section~\ref{sec:parallel} we will recall the known results about
parallelisable manifolds.  In Section~\ref{sec:flat} we will prove
that manifolds admitting flat metric connections with closed torsion
three-form are locally isometric to Lie groups admitting a
bi-invariant metric.  In Section~\ref{sec:groups} we will determine
all the ten-dimensional Lie groups with bi-invariant lorentzian
metrics, up to local isometry. We will also determine which of them is
a non-dilatonic background.  In Section~\ref{sec:supersymmetry} we
will determine the amount of supersymmetry preserved by each of these
backgrounds. Finally in Section~\ref{sec:summary} we summarise our
results.

\section{Manifolds admitting absolute parallelisms}
\label{sec:parallel}

In this section we recall the known results about parallelisable
manifolds.

A differentiable manifold $M$ is said to admit an \emph{absolute
  parallelism} if it admits a smooth trivialisation of the frame
bundle $B\to M$.  Such a trivialisation consists of a smooth global
frame and hence also trivialises the tangent bundle; whence manifolds
admitting absolute parallelisms are parallelisable in the topological
sense.  The reduction theorem for connections on principal bundles
(see, for example, \cite[Section~II.7]{KobayashiNomizu}) allows us to
think of absolute parallelisms in terms of holonomy groups of
connections.  Indeed, an absolute parallelism is equivalent to a
smooth connection on the frame bundle with trivial holonomy.  This
implies, in particular, that the connection is flat and if the
manifold is simply-connected then flatness is also sufficient.

So far these notions are purely (differential) topological and make no
mention of metrics or any other structure on the manifold.  The
question arises whether there is a metric on $M$ which is consistent
with a given absolute parallelism, so that parallel transport is an
isometry; or turning the question around, whether a given
pseudo-riemannian manifold $(M,g)$ admits a consistent absolute
parallelism.  In terms of connections, a \emph{consistent absolute
  parallelism} is equivalent to a metric connection with torsion with
trivial holonomy; or, locally, to a flat metric connection with
torsion.

Élie Cartan and Schouten \cite{CartanSchouten1,CartanSchouten2}
essentially solved the riemannian case by generalising Clifford's
parallelism on the 3-sphere in two different ways.  The three-sphere
can be understood both as the unit-norm quaternions and also as the
Lie group $\SU(2)=\Sp(1)$.  The latter characterisation generalises to
other (semi)simple Lie groups, whereas the former gives rise to the
parallelism of the 7-sphere thought of as the unit-norm octonions.  It
follows from the results of Cartan and Schouten that a
simply-connected irreducible riemannian manifold admitting a
consistent absolute parallelism (equivalently a flat metric
connection) is isometric to one of the following: the real line, a
simple Lie group with the bi-invariant metric induced from a multiple
of the Killing form, or the round 7-sphere.

Their proofs might have had gaps which were addressed by Wolf
\cite{Wolf1,Wolf2}, who also generalised these results to arbitrary
signature, subject to an algebraic curvature condition saying that the
pseudo-riemannian manifold $(M,g)$ is of ``reductive type,'' a
condition which is automatically satisfied in the riemannian case.
(See Wolf's paper for the precise condition.)  In the case of
lorentzian signature, Cahen and Parker \cite{CahenParker} showed that
one can relax the ``reductive type'' condition; completing the
classification of absolute parallelisms consistent with a lorentzian
metric.

Wolf also showed that if one also assumes that the torsion is
parallel, then, in any signature, $(M,g)$ is locally isometric to a
Lie group with a bi-invariant metric.  In fact, it is possible to show
(see below) that one obtains the same result starting with the weaker
hypothesis that the torsion three-form is closed, which is the case
needed in supergravity.

In particular, since the 7-sphere is not a Lie group, it follows that
its torsion three-form is \emph{not} closed; although it is co-closed.
This follows from the fact that the round 7-sphere $S^7$ admits
geometric Killing spinors.  Recall that a 7-manifold admits a $G_2$
structure if and only if it is spin \cite{LM}.  It follows moreover
that associated with the $G_2$ structure there is a canonical
non-vanishing spinor field $\psi$.  If $\nabla\psi=0$ then the $G_2$
structure is said to be parallel and the manifold has $G_2$ holonomy.
If $\nabla_X \psi = \lambda X \cdot \psi$ ($\lambda\neq 0$), so that
the spinor field is Killing (in the geometric sense), then the $G_2$
structure is said to be nearly parallel.  The different types of $G_2$
structures in 7-manifolds have been classified by Fernández and Gray
\cite{FernandezGray} by studying the algebraic type of $\nabla H$, or
equivalently $\nabla \psi$.  For the nearly parallel $G_2$ structures,
the torsion three-form $H$ satisfies $dH = \lambda \star H$, for some
nonzero real number $\lambda$.  (If $\lambda =0$ then the manifold has
$G_2$ holonomy and the $G_2$ structure is parallel.)  Thus, $H$ is
co-closed, but not closed.  Since $H^3(S^7)=0$, the three-form $H$
cannot be both closed and co-closed.  This means that $\AdS_3 \times
S^7$, although parallelisable, is not a \emph{parallelisable}
supergravity background.  This does \emph{not} mean that metrically
$\AdS_3 \times S^7$ cannot be a supergravity background provided we
turn on the dilaton and other fluxes, and indeed such backgrounds are
known \cite{DGT}.

The results of Cahen and Parker \cite{CahenParker} actually show that
in lorentzian signature one gets for free that the torsion is
parallel.  Therefore it follows that an indecomposable lorentzian
manifold $(M,g)$ admits a consistent absolute parallelism if and only
if it is locally isometric to a lorentzian Lie group with bi-invariant
metric.  In particular, $\AdS_7$ is \emph{not} parallelisable, even if
we do not impose any conditions on the torsion three-form, such as
that it be closed.  Therefore $\AdS_7 \times S^3$ cannot be a
parallelisable supergravity background.

It may seem surprising that the naive continuation to lorentzian
signature does not work.  One can understand this more conceptually by
realising that the consistent absolute parallelism of $S^7$ arises
from the identification of $S^7$ as the sphere of unit octonions.
There is no real division algebra, however, whose unit ``sphere'' has
lorentzian signature.  There is however a split version of the
octonions which does give rise to a consistent absolute parallelism in
a space form of signature $(3,4)$.  There is also a consistent absolute
parallelism in a complexified version of the seven-sphere
$\SO(8,\CC)/\SO(7,\CC)$.  Both these results are contained in
\cite{Wolf1,Wolf2}.

\section{Flat metric connections with closed torsion}
\label{sec:flat}

We will now show that a pseudo-riemannian manifold $(M,g)$ with a
flat metric connection with closed torsion three-form is locally
isometric to a Lie group admitting a bi-invariant metric.  This
section is based on work with Ali Chamseddine and Wafic Sabra
\cite{CFOSchiral}.

Let $(M,g)$ be a pseudo-riemannian manifold and let $D$ be a metric
connection with torsion $T$. In other words, $D g = 0$ and for all
vector fields $X,Y$ on $M$, $T:\Lambda^2TM \to TM$ is defined by
\begin{equation*}
 T(X,Y) =  D_X Y - D_Y X - [X,Y]~.
\end{equation*}
In terms of the torsion-free Levi-Civita connection $\nabla$, we have
\begin{equation*}
  D_X Y = \nabla_X Y + \half T(X,Y)~.
\end{equation*}
Since both $Dg=0$ and $\nabla g = 0$, $T$ is skew-symmetric:
\begin{equation}
  \label{eq:skew}
  g(T(X,Y),Z) = - g(T(X,Z),Y)~,
\end{equation}
for all vector fields $X,Y,Z$ and gives rise to a \emph{torsion
three-form} $H\in\Omega^3(M)$, defined by
\begin{equation*}
  H(X,Y,Z) = g(T(X,Y),Z)~.
\end{equation*}
We will \emph{assume} that $H$ is closed and in this section we will
characterise those manifolds for which $D$ is flat.

Let $R^D$ denote the curvature tensor of $D$, defined by
\begin{equation*}
  R^D(X,Y)Z = D_{[X,Y]}Z - D_X D_Y Z + D_Y
  D_X Z~.
\end{equation*}
Our strategy will be to consider the equation $R^D = 0$,
decompose it into types and solve the corresponding equations.  We
will find that $T$ is parallel with respect to both $\nabla$ and $D$,
and this will imply that $(M,g)$ is locally a Lie group with a
bi-invariant metric and $D$ the parallelising connection of Cartan and
Schouten \cite{CartanSchouten1}.

The curvature $R^D$ is given by
\begin{multline*}
  R^D(X,Y)Z = R(X,Y)Z - \half (\nabla_X T)(Y,Z) + \half (\nabla_Y
  T)(X,Z)\\
  - \tfrac14 T(X,T(Y,Z)) + \tfrac14 T(Y,T(X,Z))~,
\end{multline*}
where $R=R^\nabla$ is the curvature of the Levi-Civita connection.  The
tensor
\begin{equation*}
  R^D(X,Y,Z,W) := g(R^D(X,Y)Z,W)
\end{equation*}
takes the following form
\begin{multline*}
  R^D(X,Y,Z,W) = R(X,Y,Z,W)\\
  - \half g((\nabla_X T)(Y,Z),W) + \half g((\nabla_Y T)(X,Z),W)\\
  - \tfrac14 g(T(X,T(Y,Z)),W) + \tfrac14 g(T(Y,T(X,Z)),W)~,
\end{multline*}
where we have defined the Riemann tensor as usual:
\begin{equation*}
  R(X,Y,Z,W) := g(R(X,Y)Z,W)~.
\end{equation*}
Using equation \eqref{eq:skew} we can rewrite $R^D$ as
\begin{multline*}
  R^D(X,Y,Z,W) = R(X,Y,Z,W)\\
  - \half g((\nabla_X T)(Y,Z),W) + \half g((\nabla_Y T)(X,Z),W)\\
  + \tfrac14 g(T(X,W),T(Y,Z)) - \tfrac14 g(T(Y,W),T(X,Z))~,
\end{multline*}
which is manifestly skew-symmetric in $X,Y$ and in $Z,W$.  Observe
that unlike $R$, the torsion terms in $R^D$ do \emph{not} satisfy the
first Bianchi identity.  Therefore breaking $R^D$ into algebraic types
will give rise to more equations and will eventually allow us to
characterise the data $(M,g,T)$ for which $R^D = 0$.

Indeed, let $R^D = 0$ and consider the identity
\begin{equation*}
  \fS_{XYZ} R^D(X,Y,Z,W) = 0~,
\end{equation*}
where $\fS$ denotes signed permutations.  Since $R$ does obey the
Bianchi identity
\begin{equation*}
  \fS_{XYZ} R(X,Y,Z,W) = 0~,
\end{equation*}
we obtain the following identity
\begin{equation}
  \label{eq:bianchi}
  \fS_{XYZ} g((\nabla_X T)(Y,Z),W) =  - \tfrac12 \fS_{XYZ}
  g(T(W,X),T(Y,Z))~.
\end{equation}
Now we use the fact that the torsion three-form $H$ is closed, which
can be written as
\begin{multline*}
  g((\nabla_X T)(Y,Z),W) - g((\nabla_Y T)(X,Z),W)\\
  + g((\nabla_Z T)(X,Y),W) - g((\nabla_W T)(X,Y),Z)=0~,
\end{multline*}
or equivalently,
\begin{equation*}
  g((\nabla_W T)(X,Y),Z) = \half \fS_{XYZ} g((\nabla_X T)(Y,Z),W)~.
\end{equation*}
This turns equation \eqref{eq:bianchi} into
\begin{equation}
  \label{eq:bianchitoo}
  g((\nabla_W T)(X,Y),Z) = - \tfrac14 \fS_{XYZ} g(T(W,X),T(Y,Z))~.
\end{equation}
From this equation it follows that
\begin{equation*}
  g((\nabla_W T)(X,Y),Z) = - g((\nabla_X T)(W,Y),Z)~,
\end{equation*}
so that $g((\nabla_W T)(X,Y),Z)$ is totally skew-symmetric.  This
means that $\nabla H = dH = 0$, whence $H$ and hence $T$ are parallel.
Therefore equation \eqref{eq:bianchi} simplifies to
\begin{equation}
  \label{eq:Pluecker}
  \fS_{XYZ} g(T(W,X),T(Y,Z)) = 0~.
\end{equation}
Let us remark that $\nabla H=0$ and equation \eqref{eq:Pluecker} implies that
$D H=0$ as well.  Indeed,
\begin{footnotesize}
  \begin{align*}
    (D_W H)(X,Y,Z) &= W H(X,Y,Z) - H(D_W X,Y,Z) - H(X,D_W
    Y,Z) - H(X,Y,D_W Z)\\
    &= W H(X,Y,Z) - H(\nabla_W X,Y,Z) - H(X,\nabla_W Y,Z) -
    H(X,Y,\nabla_W Z)\\
    & \qquad - \half H(T(W,X),Y,Z) - \half H(X,T(W,Y),Z) - \half
    H(X,Y,T(W,Z))\\
    &= (\nabla_W H)(X,Y,Z) - \half \fS_{XYZ} g(T(W,X),T(Y,Z))~,
  \end{align*}
\end{footnotesize}
whence if $\nabla H=0$ and \eqref{eq:Pluecker} holds, then $D H = 0$ as
well.

Equation \eqref{eq:Pluecker} is precisely the statement that the
skew-endomorphism $\imath_W T \in \fso(TM)$ defined by $\imath_W T(X)
= T(W,X)$ leaves the torsion three-form $H$ invariant.  Indeed, the
action of $\imath_W T$ on $H$ is given by
\begin{footnotesize}
  \begin{align*}
    (\imath_W T \cdot H)(X,Y,Z) &= - H (\imath_W T(X), Y, Z) -
    H (X, \imath_W T(Y), Z) - H (X, Y, \imath_W T(Z))\\
    &= - H(T(W,X),Y,Z) - H(X, T(W,Y),Z) - H(X, Y, T(W,Z))\\
    &= - H(Y,Z,T(W,X)) + H(X, Z, T(W,Y)) - H(X, Y, T(W,Z))\\
    &= - g(T(Y,Z), T(W,X)) + g(T(X,Z), T(W,Y)) - g(T(X,Y), T(W,Z))\\
    &= - g(T(W,X), T(Y,Z)) - g(T(W,Y), T(Z,X)) - g(T(W,Z), T(X,Y))\\
    &= - \fS_{XYZ} g(T(W,X),T(Y,Z))~.
  \end{align*}
\end{footnotesize}
We pause to remark parenthetically that this shows that equation
\eqref{eq:Pluecker} is an instance of the Plücker relations in
\cite{FOPPluecker}.  More familiar, perhaps, is the fact that equation
\eqref{eq:Pluecker} is the Jacobi identity for $T$.  Indeed, notice
that
\begin{multline*}
  g(T(W,X),T(Y,Z)) = H(W,X,T(Y,Z))\\
  = H(X,T(Y,Z),W) = g(T(X,T(Y,Z)),W)~,
\end{multline*}
whence equation \eqref{eq:Pluecker} is satisfied if and only if
\begin{equation}
  \label{eq:Jacobi}
    \fS_{XYZ} T(X,T(Y,Z)) = 0~.
\end{equation}
This means that the tangent space $T_pM$ of $M$ at every point $p$
becomes a Lie algebra where the Lie bracket is given by the
restriction of $T$ to $T_pM$.  More is true and the restriction to
$T_pM$ of the metric $g$ gives rise to an (ad-)invariant scalar
product:
\begin{equation*}
  g(T(X,Y),Z) = g(X,T(Y,Z))~.
\end{equation*}

By a theorem of Wolf \cite{Wolf1,Wolf2} (based on earlier work of Élie
Cartan and Schouten \cite{CartanSchouten1,CartanSchouten2}) if $(M,g)$
is complete then it is a discrete quotient of a Lie group with a
bi-invariant metric.  In general, we can say that $(M,g)$ is locally
isometric to a Lie group with a bi-invariant metric.

Indeed, since $D$ is flat, there exists locally a parallel frame
$\{\xi_i\}$ for $TM$.  Since $\xi_i$ is parallel, from the definition
of the torsion,
\begin{equation*}
  T(\xi_i,\xi_j) = - [\xi_i,\xi_j]~.
\end{equation*}
Moreover, since $T$ is parallel relative to $D$, we see that
$[\xi_i,\xi_j]$ is also parallel with respect to $D$, whence it can be
written as a linear combination of the $\xi_i$ with constant
coefficients.  In other words, they span a real Lie algebra $\fg$.
The homomorphism $\fg \to C^\infty(M,TM)$ whose image is the
subalgebra spanned by the $\{\xi_i\}$ integrates, once we choose a
point in $M$, to a local diffeomorphism $G\to M$.  This is also an
isometry if we use on $G$ the metric induced from the one on the Lie
algebra, whence we conclude that $(M,g)$ is locally isometric to a Lie
group with a bi-invariant metric.

To conclude let us make the observation that the condition
$R^\nabla=0$ allows us to express the Riemann curvature in terms of
$T$ as follows:
\begin{equation}
  \label{eq:riemannLG}
  R(X,Y)Z = \tfrac14  T(T(X,Y),Z)~,
\end{equation}
which agrees with the standard expression for the Riemann curvature of
a bi-invariant metric on a Lie group (see, e.g.,
\cite[Ch. X, Prop.~2.12]{KobayashiNomizu}) if we identify $-T$ with the
Lie bracket, as was done above.  Contracting the above expression we
obtain an expression for the Ricci curvature which agrees with the
equation of motion for the metric coming from type II supergravity.
Furthermore, it is clear from equation \eqref{eq:action} that a
constant dilaton is consistent with its equation of motion if and only
the lagrangian vanishes.  Computing the scalar curvature from equation
\eqref{eq:riemannLG} and inserting the resulting equation into the
action \eqref{eq:action} we see that the lagrangian vanishes if and
only if the scalar curvature vanishes, or equivalently $\|H\|^2=0$.

\section{Parallelisable supergravity backgrounds}
\label{sec:groups}

Summarising the above discussion, the parallelisable supergravity
backgrounds are locally isometric to Lie groups admitting a
bi-invariant lorentzian metric with vanishing scalar curvature.
Equivalently, they are in one-to-one correspondence with
ten-dimensional Lie algebras admitting a lorentzian ad-invariant
metric and such that the structure constants satisfy $f_{abc}f^{abc} =
0$.  For a recent treatment of lorentzian Lie algebras (albeit in six
dimensions) the reader is referred to the forthcoming work
\cite{CFOSchiral}.  Here we simply summarise the result.

It follows from the structure theorem of Medina and Revoy
\cite{MedinaRevoy} (see also \cite{FSsug} for a refinement) that an
indecomposable lorentzian Lie algebra is either isomorphic to
$\fso(1,2)$ with (a multiple of) the Killing form, or else is solvable
and can be described as a \emph{double extension} $\fd_{2n+2}:=
\fd(\EE^{2n},\RR)$ of the abelian Lie algebra $\EE^{2n}$ with the
(trivially invariant) euclidean metric by a one-dimensional Lie
algebra acting on $\EE^{2n}$ via a non-degenerate skew-symmetric
linear map $J : \EE^{2n} \to \EE^{2n}$.

More concretely, the double extension $\fd_{2n+2}$ has underlying
vector space $\VV = \EE^{2(d-1)} \oplus \RR \oplus \RR$, and if
$(\bv,v^-,v^+),(\bw, w^-, w^+) \in \VV$, then their Lie bracket is
given by
\begin{equation*}
  [(\bv,v^-,v^+), (\bw, w^-, w^+)] = (v^- J(\bw) - w^- J(\bv), 0, \bv
  \cdot J(\bw))
\end{equation*}
and their inner product follows by polarisation from
\begin{equation*}
  \|(\bv,v^-,v^+)\|^2 = \bv \cdot \bv + 2 v^+ v^-~.
\end{equation*}

The unique simply-connected Lie group with Lie algebra
$\fd_{2n+2}$ is a solvable ($2n+2$)-dimensional Lie group admitting a
bi-invariant metric
\begin{equation}
  \label{eq:cwmetric}
  ds^2 = 2 dx^+ dx^- - \left<J\bx,J\bx\right> (dx^-)^2 +
  \left<d\bx,d\bx\right>~,
\end{equation}
relative to natural coordinates $(\bx, x^-, x^+)$.

Because $J$ is non-degenerate and skew-symmetric, it can always be
skew-diagonalised via an orthogonal transformation.  The
skew-eigenvalues $\lambda_1,\dots,\lambda_n$, which are different from
zero, can be arranged so that they obey: $0 < \lambda_1 \leq \lambda_2
\leq \cdots \leq \lambda_n$.  Finally a positive rescaling of $J$ can
be absorbed into reciprocal rescalings of $x^\pm$, so that we can set
$\lambda_n$, say, equal to $1$ without loss of generality.  Therefore
we see that the moduli space of metrics \eqref{eq:cwmetric} is given
by an ($n-1$)-tuple $\blambda = (\lambda_1,\ldots,\lambda_{n-1})$
where $0 < \lambda_1 \leq \cdots \leq \lambda_{n-1} \leq 1$.  We will
call such a metric $\CW_{2n+2}(\blambda)$, as they are particular
cases of Cahen--Wallach spacetimes \cite{CahenWallach}.

Before determining the ten-dimensional lorentzian Lie algebras, let us
observe that the above remarks allow us to compare with the results of
\cite[Section~4]{SSJ} on parallelisable plane waves.  First of all it
can be shown \cite{Wolf1,Wolf2} that parallelisable manifolds are
necessarily locally symmetric.  Hence parallelisable plane waves have
to be locally isometric to Cahen--Wallach spacetimes.  From the
results of Cahen and Parker in the general case or from what was
proven above for the case of closed torsion three-form, we know that
they also have to be locally isometric to Lie groups with a
bi-invariant metric.  The problem of determining which Cahen--Wallach
spacetimes are locally isometric to Lie groups with a bi-invariant
metric was solved in \cite[Section~2.3]{FSPL} and the answer consists
precisely of the metrics given above in \eqref{eq:cwmetric}, in
agreement with the results of \cite[Section~4]{SSJ}.

From the above remarks one can determine the ten-dimensional
lorentzian Lie algebras: they are either $\fso(1,2) \oplus \fg_7$ or
$\fd_{2n+2} \oplus \fg_{8-2n}$, where $\fg_d$ is a $d$-dimensional
reductive Lie algebra with a positive-definite metric.  It is easy to
come up with Table~\ref{tab:list}, where we have written the Lie
algebras and the corresponding spacetimes (up to local isometry).

\begin{table}[h!]
  \centering
  \setlength{\extrarowheight}{3pt}
  \renewcommand{\arraystretch}{1.3}
  \begin{small}
    \begin{tabular}{|>{$}l<{$}|>{$}l<{$}|}\hline
      \multicolumn{1}{|c|}{Lie algebra} & \multicolumn{1}{c|}{Spacetime}\\
      \hline\hline
      \fso(1,2) \oplus \fsu(2) \oplus \fsu(2) \oplus \RR & \AdS_3 \times
      S^3 \times S^3 \times \RR\\
      \fso(1,2) \oplus \fsu(2) \oplus \EE^4 & \AdS_3 \times S^3
      \times \RR^4\\
      \fso(1,2) \oplus \EE^7 & \AdS_3 \times \RR^7\\
      \fd_{10} & \CW_{10}(\blambda)\\
      \fd_8 \oplus \EE^2 & \CW_8(\blambda) \times \RR^2\\
      \fd_6 \oplus \fsu(2) \oplus \RR & \CW_6(\blambda) \times S^3
      \times \RR\\
      \fd_6 \oplus \EE^4 & \CW_6(\blambda) \times \RR^4\\
      \fd_4 \oplus \fsu(2) \oplus \fsu(2)  & \CW_4(\blambda) \times S^3
      \times S^3\\
      \fd_4 \oplus \fsu(2) \oplus \EE^3 & \CW_4(\blambda) \times S^3
      \times \RR^3\\
      \fd_4 \oplus \EE^6 & \CW_4(\blambda) \times \RR^6\\
      \EE^{1,1} \oplus \fsu(3) & \RR^{1,1} \times \SU(3)\\
      \EE^{1,3} \oplus \fsu(2) \oplus \fsu(2) & \RR^{1,3} \times S^3
      \times S^3\\
      \EE^{1,6} \oplus \fsu(2) & \RR^{1,6} \times S^3\\
      \EE^{1,9} & \RR^{1,9}\\ \hline
    \end{tabular}
  \end{small}
  \vspace{8pt}
  \caption{Ten-dimensional parallelisable spacetimes.}
  \label{tab:list}
\end{table}

Finally we impose the condition that the scalar curvature should
vanish, which is the consistency condition for a constant dilaton.
Since $\CW(\blambda)$ is scalar flat, any spacetime of the form
$\CW_{2n}(\blambda) \times \EE^{10-2n}$ is scalar flat.  Any
background containing an $S^3$ (or $\SU(3)$) factor can only be scalar
flat if there is a factor with negative scalar curvature to balance
it; namely $\AdS_3$.  Therefore apart from the $\CW_{2n}(\blambda)
\times \EE^{10-2n}$ backgrounds, only $\AdS_3 \times S^3 \times S^3
\times \RR$ and $\AdS_3 \times S^3 \times \RR^4$ can possibly be
consistent non-dilatonic backgrounds.  For a space form, such as
$\AdS_3$ and $S^3$, the scalar curvature is inversely proportional to
the square of the radius of curvature and the proportionality constant
only depends on the dimension, here $3$.  Therefore the scalar
curvature of the product $\AdS_3 \times S^3 \times S^3 \times \RR$ is
proportional to $-1/R_1^2 + 1/R_2^2 + 1/R_2^2$, where $R_1$ is the
radius of curvature of $\AdS_3$ and $R_2$ and $R_3$ are the radii of
curvature of the spheres.  Therefore $\AdS_3 \times S^3 \times S^3
\times \RR$ is a consistent non-dilatonic background if and only if
$1/R_1^2 = 1/R_2^2 + 1/R_2^2$.  Similary, $\AdS_3 \times S^3 \times
\RR^4$ is a consistent background if and only if the radii of
curvature of the two non-flat factors agree.

In summary, the non-dilatonic parallelisable NS-NS supergravity
backgrounds are listed in Table~\ref{tab:backgrounds}, where we have
also listed the amount of supersymmetry that is preserved.  This 
depends on the moduli.  The details of how the third column was
arrived at appear in the next section: in all cases except for one,
there is no distinction between the type IIA and type IIB theories.

\begin{table}[h!]
  \centering
  \setlength{\extrarowheight}{3pt}
  \renewcommand{\arraystretch}{1.3}
  \begin{small}
    \begin{tabular}{|>{$}l<{$}|>{$}l<{$}|l|}\hline
      \multicolumn{1}{|c|}{Lie algebra} & \multicolumn{1}{c|}{Spacetime}
      & \multicolumn{1}{c|}{Supersymmetry}\\
      \hline\hline
      \fso(1,2) \oplus \fsu(2) \oplus \fsu(2) \oplus \RR & \AdS_3 \times
      S^3 \times S^3 \times \RR & 16\\
      \fso(1,2) \oplus \fsu(2) \oplus \EE^4 & \AdS_3 \times S^3
      \times \RR^4 & 16\\
      \fd_{10} & \CW_{10}(\blambda)& 16,18(A),20,22(A),24(B),28(B)\\
      \fd_8 \oplus \EE^2 & \CW_8(\blambda) \times \RR^2& 16,20\\
      \fd_6 \oplus \EE^4 & \CW_6(\blambda) \times \RR^4& 16,24\\
      \fd_4 \oplus \EE^6 & \CW_4(\blambda) \times \RR^6& 16\\
      \EE^{1,9} & \RR^{1,9}& 32\\ \hline
    \end{tabular}
  \end{small}
  \vspace{8pt}
  \caption{Ten-dimensional non-dilatonic parallelisable NS-NS
    backgrounds.  We have adorned with (A) or (B) cases which only
    occur for type IIA or IIB, respectively.}
  \label{tab:backgrounds}
\end{table}

\section{Supersymmetry}
\label{sec:supersymmetry}

As mentioned in the introduction, the amount of supersymmetry
preserved by a non-dilatonic parallelisable supergravity background is
determined by the dilatino variation and is measured by the dimension
of the kernel of the operation of Clifford multiplication by the
torsion three-form.  Because of the bi-invariance of the metric, this
is a condition which can be analysed at the identity, whence at the
level of the Lie algebra.

Let $\fg$ be one of the above ten-dimensional lorentzian Lie algebras.
We will let $[-,-]$ and $\left<-,-\right>$ denote the Lie bracket and
the invariant metric.  Let $e_a$ be a pseudo-orthonormal frame and
define $H_{abc} := \left<[e_a,e_b],e_c\right>$.  Let $\Gamma_a$ be the
corresponding basis for the Clifford algebra $\Cl(1,9)$.  As a real
associative algebra, $\Cl(1,9) \cong \Mat(32,\RR)$ whence there is a
unique irreducible Clifford module $S$: real and of dimension $32$.
Under the spin group $\Spin(1,9)$, $S$ breaks up into $S_+ \oplus S_-$
according to chirality.  The relevant spinor representation in type
IIA is the module $S$ itself, whereas in type IIB it is the
complexification $S_+ \otimes \CC$.

We are interested in the kernel of the Clifford endomorphism $\bH =
\frac16 H_{abc} \Gamma^{abc}$ acting on the relevant module.  Notice
that $\bH$ exchanges chirality, whence $\bH: S_\pm \to S_\mp$.  Let
$\bH_\pm$ denote the restriction of $\bH$ to $S_\pm$.  We will denote
by $\bH_A$ and $\bH_B$ the relevant Clifford endomorphism in type IIA
and type IIB, respectively.  Then notice that $\dim\ker\bH_A =
\dim\ker\bH_+ + \dim\ker\bH_-$, whereas $\dim\ker\bH_B =
2\dim\ker\bH_+$, whence for type IIB, there is an even number of
supersymmetries preserved.  It is now a simple matter to scan the Lie
algebras in Table~\ref{tab:backgrounds} and examine the endomorphism
$\bH$ for each one.

Before doing so, it is convenient to write each of the different types
of Lie algebras which appear below relative to an pseudo-orthonormal
frame and compute the Clifford endomorphism $\bH$ explicitly.  In some
cases this is easily done in an explicit representation.

For example, for $\fso(1,2)$ we can choose as basis $X_0 = i\sigma_3$,
$X_1=\sigma_1$ and $X_2 =\sigma_2$, where the $\sigma_i$ are the
(hermitian) Pauli matrices.  The Lie bracket is
\begin{equation*}
  [X_0,X_1]=-2X_2 \qquad   [X_1,X_2]=2X_0 \qquad   [X_0,X_2]=2X_1
\end{equation*}
and hence $H_{012} = \left<[X_0,X_1],X_2\right> = -2$, whence the
Clifford endomorphism is 
\begin{equation*}
  \bH = -2 \Gamma^{012}~.
\end{equation*}
Now this corresponds to $\AdS_3$ with ``unit'' radius of curvature.
If we want to consider $\AdS_3$ with radius of curvature $R$ then we
have to rescale the metric $\left<-,-\right>$ by $R^2$, and the
$\Gamma^a$ matrices by $R^{-1}$, whence the Clifford endomorphism
becomes
\begin{equation*}
  \bH = -2 R^{-1} \Gamma^{012}~.
\end{equation*}

Similarly, for $\fsu(2)$, we can take as basis $X_1 = i\sigma_1$, $X_2
= i\sigma_2$ and $X_3 = i\sigma_3$.  The Lie bracket is
\begin{equation*}
  [X_1,X_2]=-2X_3 \qquad   [X_2,X_1]=-2X_1 \qquad   [X_3,X_1]=2X_2
\end{equation*}
and hence $H_{123} = \left<[X_1,X_2],X_3\right> = -2$, whence the
Clifford endomorphism, once we introduce the $S^3$ radius $R$, is
\begin{equation*}
  \bH = -2 R^{-1} \Gamma^{123}~.
\end{equation*}

Finally, we discuss the double extensions associated to
$\CW_{2n+2}(\blambda)$.  Relative to a lightcone basis $X_i,X_+,X_-$,
with $\left<X_i,X_j\right> = \delta_{ij}$ and $\left<X_+,X_-\right> =
1$, the Lie bracket is
\begin{equation*}
  [X_-,X_i] = J_{ij} X_j \qquad [X_i,X_j] = J_{ij} X_+ ~,
\end{equation*}
whence the only nonzero component of $H_{abc}$ is $H_{ij-} =
\left<[X_i,X_j],X_-\right> = J_{ij}$, whence the Clifford endomorphism
is given by
\begin{equation*}
  \bH = J_{ij} \Gamma^{ij}\Gamma^- = J_{ij} \Gamma^{ij}\Gamma_+~.
\end{equation*}

We now have all the necessary ingredients to compute the dimension of
the kernel of the Clifford endomorphism $\bH$ in each of the
parallelisable supergravity backgrounds classified in the previous
section and hence determine the amount of supersymmetry which they
preserve.

\subsection{$\AdS_3 \times S^3 \times S^3 \times \RR$}

Introducing radii $R_1$, $R_2$ and $R_3$ for the $\AdS_3$, and the two
3-spheres respectively, the Clifford endomorphism is (up to an overall
scale) given by
\begin{equation*} 
  \bH = R^{-1}_1 \Gamma^{012} + R^{-1}_2 \Gamma^{345} + R^{-1}_3
  \Gamma^{678}~.
\end{equation*}
This endomorphism is invertible unless $R_1^{-2} = R_2^{-2} +
R_3^{-2}$, which is precisely the consistency condition for a constant
dilaton.\footnote{This is not a coincidence.  If the Clifford
  endomorphism $\bH$ has kernel, so does its square.  Using the Jacobi
  indentity, $\bH^2 = \|H\|^2 \1$, whence supersymmetry implies that
  $\|H\|^2 = 0$ or, equivalently, that the scalar curvature vanishes.}
In this case the kernel is sixteen-dimensional.  The generator
$\Gamma_9$ anticommutes with $\bH$ and hence preserves the kernel of
$\bH$.  Since $\Gamma_9$ is invertible and exchanges chirality, we see
that $\dim\ker\bH_+ = \dim\ker\bH_- = 8$.  Therefore $\dim\ker\bH_A =
\dim\ker\bH_B = 16$ and this is a half-BPS background for both type
IIA and type IIB supergravity.  This is in agreement with the
supergravity results of \cite{CT,BPS,GMT} and from conformal field
theory in \cite{EFGT}.

\subsection{$\AdS_3 \times S^3 \times \RR^4$}

This is the limit $R_3\to \infty$ of the previous example.  Therefore
$R_1 = R_2$ and the background preserves 16 supersymmetries both for
type IIA and IIB \cite{Malda}.

\subsection{$\CW_{10}(\blambda)$}

The Clifford endomorphism in this case takes the form
\begin{equation*}
  \bH = (\lambda_1 \Gamma^{12} + \lambda_2 \Gamma^{34} + \lambda_3
  \Gamma^{56} + \lambda_4 \Gamma^{78}) \Gamma_+~,
\end{equation*}
where we have reintroduced the scale ($\lambda_4$) for convenience.
Because of the $\Gamma_+$, such endomorphism has kernel and in fact,
the dimension of the kernel is at least 16, both for type IIA and type
IIB.  The kernel may be larger, however, depending on whether the
endomorphism
\begin{equation*}
  \bJ = \lambda_1 \Gamma^{12} + \lambda_2 \Gamma^{34} + \lambda_3
  \Gamma^{56} + \lambda_4 \Gamma^{78}
\end{equation*}
has any kernel in the subspace $\ker\Gamma_-$ of the relevant Clifford
module.

This question can be analysed group-theoretically (see, for example,
\cite[Section~2.2]{FigSimBranes} for a more detailed analysis in a
related problem) by interpreting $\bJ$ as an element of the Cartan
subalgebra of $\fso(8) \subset \fso(1,9)$ acting on the
$16$-dimensional representation $\ker\Gamma_-$ (for type IIA) or on
the complexification of the $8$-dimensional representation
$\ker\Gamma_- \cap S_+$ (for type IIB).

In terms of $\fso(8)$ irreducible representations, $\Delta:=
\ker\Gamma_- = \Delta_+ \oplus \Delta_-$, where $\Delta_\pm$ are the
half-spin representations: $\repre{8}_s$ and $\repre{8}_c$,
respectively.  The weights of $\Delta$ with respect to the above basis
for the Cartan subalgebra of $\fso(8)$ are $(\pm1,\pm1,\pm1,\pm1)$
where the signs are uncorrelated, for a total of $2^4 = 16$ weights.
The weights for which the products of the signs is $\pm 1$ correspond
to $\Delta_\pm$.

Let us first of all consider the case of type IIA. The action of $\bJ$
on the representation $\Delta$ is diagonal with eigenvalues $\pm
\lambda_1 \pm \lambda_2 \pm \lambda_3 \pm \lambda_4$. Setting each of
these expressions to zero gives rise to 8 hyperplanes in the
four-dimensional parameter space of the $\blambda =
(\lambda_1,\lambda_2,\lambda_3,\lambda_4)$.  For $\blambda$ away from
such hyperplanes, the solution preserves no extra
supersymmetry. However for $\blambda$ in the union of the hyperplanes,
the solution preserves extra supersymmetry---how much depending on to
how many of these hyperplanes $\blambda$ belongs.

If $\blambda$ belongs to one and only one hyperplane, then there are
two extra supersymmetries, for a total of $18$.  If $\blambda$ belongs
to the intersection of two (but not more) hyperplanes and that, as
should be the case for $\CW_{10}$, no $\lambda_i$ should vanish, then
there are two more for a total of $20$.  If $\blambda$ belongs to the
intersection of three (but not more) hyperplanes and again with no
$\lambda_i$ vanishing, there are two more zero eigenvalues for a total
of $22$.  There are no $\blambda$ in the intersection of four
hyperplanes with all $\blambda_i$ nonzero.

Let us now consider type IIB.  We must restrict ourselves to
the four hyperplanes for which the products of the signs is positive
and count them with multiplicity four.  If $\blambda$ lies in
precisely one of the hyperplanes, then there are an additional $4$
supersymmetries for a total of $20$.  If $\blambda$ lies in the
intersection of precisely two hyperplanes, but again with no
$\lambda_i$ vanishing, then there are an additional $4$ for a total of
$24$.  If $\blambda$ lies in the intersection of precisely three
hyperplanes (and again no $\lambda_i$ vanishing) then there are an
additional $4$ for a total of $28$ supersymmetries.  There are no
nonzero $\blambda$ in the intersection of all four hyperplanes.

\subsection{$\CW_8(\blambda) \times \RR^2$}

This corresponds to $\lambda_4=0$ in the previous case, say.  There is
no distinction here between type IIA and type IIB, because $\Gamma^8$,
say, exchanges chirality and anticommutes with $\bH$.

Generically there will not be further supersymmetries; but if
$\blambda=(\lambda_1, \lambda_2, \lambda_3)$ lies in the union of the
four hyperplanes $\lambda_1 \pm \lambda_2 \pm \lambda_3 = 0$
(uncorrelated signs) then there will be supersymmetry enhancement.  If
$\blambda$ lies in one and only one hyperplane there will be an extra
$4$ supersymmetries, for a total of $20$.  Any $\blambda$ in the
intersection of any two of these hyperplanes automatically has some
$\lambda_i = 0$, whence it does not correspond to this background.

\subsection{$\CW_6(\blambda) \times \RR^4$}

This case corresponds to putting $\lambda_3=0$ in the above case.  Now
for generic $\blambda=(\lambda_1,\lambda_2)$ there are no extra
supersymmetries, but for $\blambda$ in the union of the two
hyperplanes $\lambda_1 \pm \lambda_2 = 0$ there is supersymmetry
enhancement.  The only allowed possibility (since $\blambda\neq0$) is
that it lies in precisely one of hyperplanes.  In this case there are
$8$ extra supersymmetries for a total of $24$.

\subsection{$\CW_4(\blambda) \times \RR^6$}

The Clifford endomorphism is
\begin{equation*}
  \bH = \lambda_1 \Gamma^{12} \Gamma_+~,
\end{equation*}
which clearly has no extra supersymmetries than those in the kernel of
$\Gamma_+$.  Hence this backgrounds preserves $16$ supersymmetries.

\subsection{$\RR^{1,9}$}

This is of course the flat vacuum solution which preserves all
supersymmetries.

\section{Summary}
\label{sec:summary}

In summary, we have classified (up to local isometry) all the
non-dilatonic parallelisable NS-NS backgrounds of ten-dimensional type
II supergravity.  Parallelisability implies that the geometry is that
of one of the parallelised lorentzian Lie groups listed in
Table~\ref{tab:list} and demanding that a constant dilaton obeys its
equation of motion reduces the possibilities further to those in
Table~\ref{tab:backgrounds}.  We have moreover shown that they
preserve $16,18,20,22,24$ or $32$ supersymmetries for type IIA, and
$16,20,24,28$ or $32$ for type IIB.  The cases with geometry $\CW_{2n}
\times \EE^{10-2n}$ are symmetric plane waves and our results agree
with those in \cite{SSJ}.  Since symmetric plane waves are in
particular homogeneous, all the plane waves in the tables are
contained in the classification of homogeneous plane waves of Blau and
O'Loughlin \cite{BOLhpw}.  Some of the plane waves in the tables have
also appeared in \cite{Michelson,CLPppM,ChrisJerome,BenaRoiban}.

Being lorentzian Lie groups admitting a bi-invariant metric and having
no other fields turned on but the NS-NS three-form, string propagation
on these parallelisable backgrounds is described by a WZW model and
hence amenable to standard techniques in conformal field theory.  In
particular it is possible to determine the symmetric D-branes for all
the backgrounds in Table~\ref{tab:backgrounds} using the techniques of
\cite{AS,FFFS,SDNotes} and indeed for many of these backgrounds this
has already been done \cite{Sads3,FSNW,FSS3,BPAdS2,FSPL}.

The spacetimes in Table~\ref{tab:backgrounds} are related by three
types of limits: large radius limits which in essence flatten
different factors in the metric, degenerations of the Cahen--Wallach
metrics by taking some of the eigenvalues $\lambda_i$ to zero, and
Penrose--Güven limits \cite{PenrosePlaneWave,GuevenPlaneWave}.

As was explained in \cite{FSPL} for $\AdS_3 \times S^3 \to
\CW_4(\blambda)$, but the idea clearly generalises, the Penrose limits
can be understood as group contractions.  Indeed, suppose that $\gamma
\subset G$ is a null geodesic.  It is determined uniquely by its
initial point $\gamma(0)$ and its initial direction $[\dot\gamma(0)]$
is the celestial sphere at $T_{\gamma(0)}G$.  From the covariance
property of \cite[Section~2.4]{Limits} we can apply an isometry to
$\gamma$ without changing the Penrose limit (up to isometry).  Using
left-translations, say, we can take $\gamma(0)$ to be the identity.
Then $\dot\gamma(0)$ is a null vector in the Lie algebra, which
generates a one-parameter subgroup $H \subset G$.  Since the metric on
$G$ is bi-invariant, one-parameter subgroups are geodesics, whence
$\gamma = H$.  Then as shown for a particular example in \cite{FSPL}
(see also \cite{ORS}) the Penrose limit along $H$ is the Inönü--Wigner
contraction of $G$ along $H$.

It is also possible to argue in what superficially appears to be more
generality, that parallelisability is a hereditary property of the
Penrose limit.  This is because the condition of
parallelisability can be phrased in terms of the existence of parallel
sections in a bundle with connection.  As discussed in \cite{Limits},
extending the results in \cite{Geroch}, parallel sections are
preserved in the Penrose limit, hence if there exists a parallel frame
before the limit there continues to be one afterwards.  There is no
need for such a general argument, though, as the results in this paper
and in \cite{CahenParker} allows us to limit ourselves to Lie groups
with bi-invariant metrics, a class of spaces preserved by group
contractions.

Finally let me remark that to complete the classification of
parallelisable NS-NS backgrounds there remains to study the
possibility of turning on the dilaton.  As explained briefly in the
introduction, there are two conditions to impose: first the equation
of motion of the dilaton itself and secondly the gradient is
constrained to lie along central directions of the Lie algebra.  Given
the explicit structure of the Lie algebra it is only a matter of
patience to determine the possibilities.

\section*{Acknowledgments}

It is a pleasure to thank Ali Chamseddine and Wafic Sabra for
discussions and for collaboration on related issues, and Matthias Blau
for correspondence and encouragement.  I would also like to thank Leo
Pando-Zayas, Darius Sadri and Shahin Sheikh-Jabbari for correspondence
about their work on parallelisable supergravity solutions and for
spotting a careless error in an earlier version; and Michael Duff for
his comments on an earlier version.  Their comments have improved the
present paper.

This work was done during a visit to the School of Natural Science of
the IAS, whom I would like to thank for the hospitality and support
and for providing such a pleasant environment.  In particular, I am
grateful to Juan Maldacena for the invitation.  This research is
partially funded by the EPSRC grant GR/R62694/01.  I am also a member
of EDGE, Research Training Network HPRN-CT-2000-00101, supported by
The European Human Potential Programme.

\bibliographystyle{amsplain}
\bibliography{AdS,AdS3,ESYM,Sugra,Geometry,CaliGeo}

\end{document}